
 \input phyzzx
 \PHYSREV
 \doublespace
 \unnumberedchapters
 \pubnum{5892\cr CERN-TH.6620/92 \cr}
 \pubtype{T/E}
 \titlepage
 \title{ On Testing V-A in $\Lambda_b$ Decays
 \doeack}
 \author{Michael Gronau\foot{On leave of absence from the Physics
 Department, Technion, Haifa, Israel.}}
 \SLAC
 \andauthor{Seiichi Wakaizumi\foot{Permanent address: Department of
 Physics, School of Medical Sciences, University of Tokushima,
 Tokushima, Japan.}}
 \address{ \break CERN, Theory Division \break CH-1211 Geneva 23,
 Switzerland}
 \abstract
We comment on a recent suggestion by Amundson, Rosner, Worah and Wise
to test the chirality of the $b$-quark
decay coupling via polarized $\Lambda_b$ baryons produced in
$e^+e^-\rightarrow Z\rightarrow \Lambda_b +X$. We study the effect of
contributions from an amplitude in which a right-handed $b$ to $c$
current couples to a V-A lepton current.

 \submit{Physical Review D}
 \endpage


$b$-quark decays are known to be governed by very weak couplings,
which in the Standard Model are given by the two tiny mixing
angles $\theta_{23}$,~ $\theta_{13}$%
\REF\pdbook{Particle Data Group,
Phys. Lett. {\bf B239}, 1 (1990).}\refend .
These decays are therefore
very sensitive to new kinds of interactions, and in particular to right-
handed couplings%
\REF\bwork{M. Gronau, talk at the Conference on B Factories: The State
of the Art in Accelerators, Detectors and Physics, Stanford, CA, April
6-10, 1992, ed. D. Hitlin, to be published.}\refend .
Recently we described a viable $SU(2)_L\times SU(2)_R\times U(1)$
model, in which $b$-quarks decay purely right-handedly%
\REF\rightb{M. Gronau and S. Wakaizumi, Phys. Rev. Lett. {\bf 68}, 1814
(1992); M. Gronau and S. Wakaizumi, in B Decays, ed. S. Stone, (World
Scientific, Singapore, 1992) p. 479; M. Gronau, Phys. Lett. {\bf B288},
90 (1992).}\refend . We pointed out that measurement of
the lepton asymmetry in
$B\rightarrow D^* \ell\nu$%
\REF\asym{CLEO Collaboration, S. Sanghera {\sl et al.},
The CLEO Collaboration, Cornell University
Report No. CNLS 92/1156 (1992), to be published.}\refend
cannot distinguish our model from the Standard Model, since in this
process the lepton current in our model is dominantly
V+A%
\REF\chirality{M.Gronau and S. Wakaizumi, Phys. Lett. {\bf B280}, 79
(1992).}\refend .
A nice test, which may distinguish between the exchange of left- and
right-handed gauge bosons in $b$ decays was suggested by Amundson,
Rosner, Worah and Wise%
\REF\rosner{J.F. Amundson, J.L. Rosner, M. Worah and M.B. Wise,
University of Chicago Report No. EFI 92-38-Rev, 1992, to be published.}
\refend .
These authors noted that the electron spectrum from
highly polarized $\Lambda_b$'s produced in $e^+e^-\rightarrow
Z\rightarrow\Lambda_b +X$ is significantly harder for V-A than for
V+A quark and lepton currents. Thus, ongoing experiments at LEP, in
which leptons from $\Lambda_b$ decay were already observed%
\REF\Zdecay{ALEPH Collaboration, D. Decamp {\sl et al.}, Phys. Lett.
{\bf B278}, 209 (1992); OPAL Collaboration, P.D. Acton {\sl et al.},
Phys. Lett. {\bf B281}, 394 (1992).}\refend\ ,
may offer an early test of the model.

In this brief note we wish to study the effect of $W_L-W_R$ mixing,
which exists in our model in addition to $W_R$ exchange%
\refmark\rightb\ ,
to see how much it can affect the electron spectrum
calculated in \refmark\rosner\ .
Also, as a model-independent study and
to demonstrate another version of right-handed $b$ to $c$ couplings, we
will first calculate the electron spectrum for a V+A quark coupling,
assuming
that the lepton current is purely left-handed. Such a possibility is
outside the parameter range of the model of \refmark\rightb\ ,
since it corresponds to decays due to $W_L-W_R$ mixing alone.
This case, which seems to be one's first guess of what
right-handed $b$ couplings might be, was
recently excluded by the $B\rightarrow D^* \ell\nu$ data\refmark\asym\ .
Our purpose of including a discussion of this case is to
show that also in the case of $\Lambda_b$ decays it
can be most easily distinguished from other cases studied here.

We use the physics of the heavy quark symmetry presented in\refmark
\rosner%
\REF\korner{See also J.G. Koerner and M. Kraemer, Phys. Lett.
{\bf B275}, 475 (1992); T. Mannel and G.A. Schuler, Phys.
Lett. {\bf B279}, 194 (1992).}\refend
and the notations of\refmark\rosner\
to describe the lepton spectrum in terms of the free-quark decay
$b\rightarrow c e^-\overline{\nu}_e$.
For a V+A $b$ to $c$ coupling and a V-A lepton current the normalized
electron decay distribution in the $b$ rest frame is given by%
\REF\tsai{Y.S. Tsai, Phys. Rev. {\bf D19}, 2809 (1979).}\refend :
$$
{1\over\Gamma}{d^2\Gamma\over dx d(\cos\psi)}={6x^2(1-\zeta)^2\over
f(m^2_c/m^2_b)}(1-x)(1-{\sl P}\cos\psi)~,\eqno\eq
$$
$$
x\equiv 2E^*/m_b~,~~~~~~~~\zeta\equiv m^2_c/[m^2_b(1-x)]~.
$$
$f(y)$ is a well-known phase-space function, and $E^*,~\psi$ are the
energy of the electron and its angle with respect to
the $b$-quark polarization. The polarization is almost complete,
${\sl P}=-0.93$.
The boost from the $b$ rest-frame to the Z frame, in which $Z\rightarrow
b{\overline b}$ occurs, is described in\refmark\rosner\ .
For ${\sl P}=-1$ we find from Eq.(1) the electron
energy spectrum shown in Fig. 1(c). We used the values of $m_b=5~{\rm
GeV},~m_c=1.66~{\rm GeV},~E_b=45~{\rm GeV}$ from \refmark\rosner\ ,
and chose a minimum electron transverse momentum of $p^{min}_T=0.8~{\rm
GeV}/c$. This spectrum should be compared with the
two spectra of \refmark\rosner\ using the same momentum cut, shown in
Fig.1(a), Fig.1(b), which describe the cases of V-A and V+A
quark and lepton currents, respectively. The difference is striking.
The distribution of Fig.1(c) peaks at a considerably higher energy
($14~{\rm GeV}$ instead of $7-9~{\rm GeV}$) and involves many fewer
low energy electrons than the two other distributions.

The possibility of a V+A $b$ to $c$ current coupled to a V-A lepton
current was recently excluded by measurement of the forward-backward
asymmetry in $B\rightarrow D^* \ell\nu$\refmark\asym\ . This measurement
favors the two cases in which both quark and lepton currents are either
V-A or V+A\refmark\chirality\ . The first case corresponds to the
Standard Model, while the second one describes the dominant $W_R$
exchange contribution in model\refmark\rightb\ .
For both cases the measured lepton angular distribution
sets $95\%$ C.L. upper limits on the
allowed rates coming from amplitudes in which the quarks and leptons
couple with opposite chiralities.
The form-factor-dependent limits on the ratio of rates of opposite
and equal chiralities are at the level of
$30\%$. The implication of these limits on the model \refmark\rightb\
is a bound on the $W_L-W_R$ mixing parameter, $\zeta_g$:
$$
\big({\zeta_g\over\beta_g}\big)^2<0.30~,~~~~~~~~~\beta_g\equiv{g^2_R
\over g^2_L}{M^2_L\over M^2_R}~.\eqno\eq
$$

We use this constraint to study within model \refmark\rightb\ the effect
of $W_L-W_R$ mixing on the lepton energy spectrum of $\Lambda_b$
decay. Fig.1(d) describes the spectrum corresponding to
$(\zeta_g/\beta_g)^2=0.29$, which is just below the limit (2). The peak
of this distribution lies between the peaks of the V-A and V+A
distributions, Fig.1(a) and Fig.1(b), respectively. That is,
the effect of $W_L-W_R$ mixing is to diminish the difference between
the distributions of our model and the Standard Model. Nevertheless,
the feature of a considerably lower high-energy electron tail persists
in our model. An observation of a higher tail, as in Fig.1(a),
would clearly favor the Standard Model.

We thank J. Amundson, J. Rosner and Y.S. Tsai for useful discussions.
One of us (S. W.) is grateful to R. Kleiss for his help in
using the CERN computer system.
\endpage

\refout
\endpage
\centerline{\bf FIGURE CAPTION}
\item{\rm FIG.1.} Distributions in electron laboratory energy for
inclusive semileptonic $\Lambda_b$ decays from $e^+e^-\rightarrow Z
\rightarrow \Lambda_b +X$, with $p^{min}_T=0.8~{\rm GeV}/c$.
(a) Standard Model \refmark\rosner\ , (b) $W_R$ exchange \refmark\rosner\
, (c) V+A quark coupling and V-A lepton coupling, (d) Model \refmark
\rightb\ with $(\zeta_g/\beta_g)^2=0.29$.

 \bye